\documentclass[aps,pra,reprint,superscriptaddress,showpacs]{revtex4-1}
\usepackage{graphicx,amsmath,amssymb,ulem}
\usepackage{verbatim}
\usepackage{ragged2e}
\usepackage{color}

\begin{document}
\title{ Bright squeezed vacuum in a nonlinear interferometer: frequency/temporal Schmidt-mode description.}
\author{P.~R.~Sharapova}
\affiliation{Department of Physics and CeOPP, University of Paderborn,
Warburger Strasse 100, D-33098 Paderborn, Germany}
\author{O.~V.~Tikhonova}
\affiliation{Physics Department, Moscow State University,  Leninskiye Gory 1-2, Moscow 119991, Russia}
\affiliation{Skobeltsyn Institute of Nuclear Physics, Lomonosov Moscow State University, Moscow 119234, Russia}
\author{S.~Lemieux}
\affiliation{Department of Physics, University of Ottawa, 25 Templeton Street, Ottawa, Ontario K1N 6N5, Canada}
\author{R.~W.~Boyd}
\affiliation{Department of Physics, University of Ottawa, 25 Templeton Street, Ottawa, Ontario K1N 6N5, Canada}
\affiliation{Institute of Optics, University of Rochester, Rochester, New York 14627, USA}
\author{M.~V.~Chekhova}
\affiliation{Physics Department, Moscow State University, Leninskiye Gory 1-2, Moscow 119991, Russia}
\affiliation{Max-Planck Institute for the Science of Light, \\  Staudtstrasse 2, Erlangen  D-91058, Germany}
\affiliation{University of Erlangen-N\"urnberg, Staudtstrasse 7-B2, Erlangen  D-91058, Germany}

\begin{abstract}
Control over the spectral properties of the bright squeezed vacuum (BSV), a highly multimode non-classical macroscopic state of light that can be generated through high-gain parametric down conversion, is crucial for many applications. In particular, in several recent experiments BSV is generated in a strongly pumped SU(1,1) interferometer to achieve phase supersensitivity, perform broadband homodyne detection, or tailor the frequency spectrum  of squeezed light. In this work, we present an analytical approach to the theoretical description of BSV in the frequency domain based on the  Bloch-Messiah reduction and the Schmidt-mode formalism. As a special case we consider a strongly pumped SU(1,1) interferometer. We show that different moments of the radiation at its output depend on the phase, dispersion and the parametric gain in a nontrivial way, thereby providing additional insights on the capabilities of nonlinear interferometers. In particular, a dramatic change in the spectrum occurs as the parametric gain increases.
\end{abstract}
\maketitle

\section{Introduction}
Bright squeezed vacuum is a macroscopic non-classical state of light that exhibits strong correlations between the signal and idler beams (twin-beam squeezing)~\cite{Jedr, Bondani, Brida, Agafonov}, quadrature squeezing ~\cite{Lett,Shaked2017}, polarization entanglement~\cite{nonsep} etc., making it attractive for applications in quantum imaging~\cite{Brida2010,Boyer,Bondani_narrow} and quantum metrology~\cite{Brida_OpEx2010,Manceau}. The BSV produced in a traveling-wave parametric amplifier is characterized by a highly multimode structure  \cite{Boyer, Boyer1, OptLett}. Depending on the choice of modes, one can observe quadrature squeezing or twin-beam squeezing - two different nonclassical effects \cite{Braunstein}. Due to the strong multiphoton correlations and complicated mode structure, the theoretical description of BSV is difficult.

Earlier works on the theoretical description of BSV \cite{Boyd,Wasilewski,Dayan,Christ,Eckstein} used a numerical approach based on solving a set of coupled integro-differential equations. In Refs.~\cite{Boyd,Wasilewski,Dayan}, the Heisenberg picture was used in the Fourier space, and the analytical solution was only found for a very narrowband pump. In Ref.~\cite{Christ,Eckstein}, broadband (Schmidt) modes were introduced and the effect of time-ordering was considered, followed by a numerical treatment.  However, recent experiments where the spectral properties of BSV are modified, in particular by using it in a nonlinear interferometer~\cite{Vered,Shaked,altern,exp}, are still lacking a detailed theoretical description.

Here we present a consistent analytical approach to the description of BSV in the frequency domain. Our approach is based on the collective Schmidt operators and allows us to take into account multiphoton correlations and nonclassical features of BSV radiation and to analyze different characteristics of BSV for various experimental configurations.  In particular, we analyse a nonlinear SU(1,1) interferometer~\cite{Yurke, Jing, Hudelist, Klyshkoint, Vered, Shaked} containing a dispersive medium~\cite{exp,altern}, which allows one to engineer the spectral properties of BSV. High-gain effects, such as the dramatic narrowing of the BSV spectrum and the generation of tunable two-color BSV, as well as the transition from low to high parametric gain, are described in terms of the Schmidt modes.  The basic idea of the developed theoretical approach appears to be rather general and can be used to describe the spatial properties of BSV as well~\cite{PRA}.

\section{FORMALISM OF THE FREQUENCY SCHMIDT MODES}
Parametric down-conversion~(PDC) in a crystal with a quadratic susceptibility $\chi^{(2)}(\mathbf{r})$ is described by the following Hamiltonian \cite{Klyshkoint} in terms of electromagnetic field operators:
\begin{equation}
H\sim\int d^3\mathbf{r}\chi^{(2)}(\mathbf{r})E_p^{(+)}(\mathbf{r},t)E_s^{(-)}(\mathbf{r},t)E_i^{(-)}(\mathbf{r},t)+\textrm{h.c.},
\label{Ham}
\end{equation}
where $s,i,p$ indices correspond to the signal, idler, and pump radiation, respectively. In this work, in contrast to Ref.~\cite{PRA}, we consider a pulsed pump, for which the envelope of $E_p^{(+)}(\mathbf{r},t)$ depends on time.

We assume a classical pump with a Gaussian envelope, $E_p^{(+)} (\mathbf{r},t)= E_0 e^{-\frac{t^2}{2\tau^2}}e^{i( \mathbf{k}_p \mathbf{r}-\omega_p t)}$,  with the full width at half maximum (FWHM) of the intensity pulse being $2\sqrt{\ln 2}\tau$. First, we will consider the case of a single crystal where PDC is produced.  Further, this model will be generalized to  other experimental configurations. By using the  expansion of the quantum fields over plane-wave modes, $E_{s,i}^{(-)} (\mathbf{r},t)=\int d{\mathbf{k}_{s,i}} C_{\mathbf{k}_{s,i}} e^{-i( \mathbf{k}_{s,i} \mathbf{r}-\omega_{s,i} t)} a^\dagger_{\mathbf{k}_{s,i}}$, with the summation replaced for convenience by integration, the Hamiltonian becomes
\begin{eqnarray}
H\sim i \iint d \mathbf{k}_s  d \mathbf{k}_i d^3\mathbf{r} \chi^{(2)}(\mathbf{r}) E_0  C_{\mathbf{k}_{s}}C_{\mathbf{k}_{i}} e^{-\frac{t^2}{2\tau^2}}\times
\nonumber\\
e^{i (\mathbf{k}_p-
\mathbf{k}_s-\mathbf{k}_i) \mathbf{r}}
e^{i (\omega_s+\omega_i-\omega_p)t}a^\dagger_{\mathbf{k}_s}a^\dagger_{\mathbf{k}_i} +\textrm{h.c.}
\label{Ham1}
\end{eqnarray}
In the Hamiltonian (\ref{Ham1}) the integration runs over all wavevectors of the signal and idler photons and over  the three spatial variables. However, in what follows, we will restrict our consideration to the collinear propagation of the photons only, and neglect the transverse wavevector components. Then, the integration over the three-dimensional wavevector domain is equivalent to the integration over frequencies. From the experimental viewpoint it means, in particular, that we consider a sufficiently broad spatial pump beam.  In this case, the Hamiltonian can be written as
\begin{eqnarray}
H\sim i \iint d \omega_s  d \omega_i \int_{-L}^{0} d z e^{-\frac{t^2}{2\tau^2}}e^{i (k_p-k_s-k_i) z}\times
\nonumber\\
e^{i (\omega_s+\omega_i-\omega_p)t}a^\dagger_{\omega_s}a^\dagger_{\omega_i} +\textrm{h.c.},
\label{Ham2}
\end{eqnarray}
where we assume that coefficients $C_{\mathbf{k}_{s,i}}$ are frequency independent, and  that $\chi^{(2)}$ is constant over the length of the crystal $L$ $(\chi^{(2)}_0)$ and zero elsewhere.

Let us represent the Gaussian temporal envelope of the pump as a Fourier transform: $\exp\{-\frac{t^2}{2\tau^2}\}\exp\{-i\omega_p t\}=\int d\omega \exp\{-\frac{(\omega-\omega_p)^2}{2 \Omega^2}\}\exp\{-i\omega t\}$, where $\Omega=1/\tau$. Then, we assume that each photon of the pump spectrum gives rise to the signal and idler photons with the energy mismatch being exactly zero, $\omega=\omega_s+\omega_i$. This approximation is well satisfied for an SU(1,1) interferometer because of the effective narrowing of the spectrum due to the nonlinear interference~\cite{Klyshkoint}. This leads to a delta-function $\delta (\omega-\omega_s-\omega_i)$ removing the integration over $\omega$, and the Hamiltonian takes the form
\begin{eqnarray}
H=i\hbar\Gamma \iint d \omega_s  d \omega_i \exp\{-\frac{(\omega_s+\omega_i - \omega_p)^2}{2 \Omega^2}\} \times
\nonumber\\
\ \ \int_{-L}^{0} d z e^{i (k_p-k_s-k_i) z} a^\dagger_{\omega_s}a^\dagger_{\omega_i}, \ \
\label{Ham21}
\end{eqnarray}
where  $a^\dagger_{\omega_{s,i}}$ are the photon creation operators for the monochromatic signal (idler) frequency modes, and $\Gamma\sim\chi^{(2)}_0 E_0$ is the effective coupling  strength.

The approximation of zero energy mismatch restricts the generality of our model, which, for instance, fails to describe the broadening of the spectrum with increasing parametric gain~\cite{Spasibko2012}. On the other hand, we take into account the whole spectral width of the pump, signal/idler pulses and the wavevector mismatch, which allows us to obtain the analytical solution to the problem and to describe many features of BSV. After integrating in $z$, the PDC Hamiltonian can be represented in the simple form
\begin{equation}
H=i\hbar\Gamma\int d \omega_s  d \omega_i F(\omega_s,\omega_i)a^\dagger_{\omega_s}a^\dagger_{\omega_i}+\textrm{h.c.},
\label{Ham3}
\end{equation}
with the two-photon amplitude (TPA) $F(\omega_s, \omega_i)$ depending only on the signal and idler frequencies,
\begin{eqnarray}
F(\omega_s,\omega_i)=C \exp\{-\frac{(\omega_s+\omega_i-\omega_p)^2}{2 \Omega^2 }\}\mathrm{sinc}(\frac{\Delta k L}{2})\times
\nonumber\\
\times\exp\{-i\frac{\Delta k L}{2} \}, \ \ \ \ \ \
\label{TPA}
\end{eqnarray}
where $C$ is the normalization constant and $\Delta=k_p(\omega_s+\omega_i)-k_s(\omega_s)-k_i(\omega_i)$ is the wavevector mismatch inside the crystal. The signal and idler wave vectors $k_{s,i}=n_{s,i}(\omega_{s,i})\omega_{s,i}/c$ depend on the refractive indices $n_{s,i}(\omega_{s,i})$ which can be calculated by using dispersion (Sellmeier) formulas \cite{Sellm}.
The interaction Hamiltonian (\ref{Ham3}) indicates correlations between the signal and idler monochromatic-wave photons. It is more convenient to introduce new spectral modes that will be independent of each other. Such a procedure is  similar to using normal coordinates for the description of interacting harmonic oscillators. In our case we can use the Schmidt decomposition \cite{Law} and present a bipartite TPA as
\begin{equation}
F(\omega_s,\omega_i)=\sum_{n} \sqrt{\lambda_{n}}u_{n}(\omega_s)v_{n}(\omega_i),
\label{Schmidt}
\end{equation}
where $\lambda_{n}$ are the eigenvalues and $u_{n}(\omega_s),v_{n}(\omega_i)$ are the Schmidt modes \cite{Fedorov, FedorovPRA}.

After the Schmidt decomposition, new photon operators can be introduced that are responsible for the creation/annihilation of a photon not with a certain frequency but with the spectral distribution determined by a certain Schmidt mode function:
\begin{eqnarray}
A_{n}^\dagger=\int d\omega_s u_{n}(\omega_s)a^\dagger_{\mathbf{\omega}_s},
\nonumber\\
B_{n}^\dagger=\int d\omega_i v_{n}(\omega_i)a^\dagger_{\mathbf{\omega}_i}.
\label{ops}
\end{eqnarray}
 The Schmidt-mode operators (\ref{ops}) are similar to the broadband operators used in \cite{Christ}. In terms of these operators, the PDC Hamiltonian is diagonalized~\cite{Braunstein},
\begin{equation}
H=i\hbar\Gamma \sum_{n} \sqrt{\lambda_{n}}(A_{n}^\dagger B_{n}^\dagger-A_{n} B_{n}).
\label{Hams}
\end{equation}
The new modes are independent and the operators (\ref{ops}) satisfy the usual commutation relations
\begin{equation}
[ A_n,A_m^\dagger]=\delta_{mn} ,\,\,
[ A_n,B_m^\dagger]=0.
\label{comm}
\end{equation}
Using the commutation relations one can obtain the Heisenberg equations for the Schmidt modes,
\begin{eqnarray}
\frac{d A_{n}}{dt}=\Gamma \sqrt{\lambda_{n}}B_{n}^\dagger , \ \
\nonumber\\
\frac{d B_{n}^\dagger}{dt}=\Gamma \sqrt{\lambda_{n}}A_{n}.  \ \ \
\label{Heis}
\end{eqnarray}
The solutions to these equations are given by the Bogolyubov transformations and provide the output operators, after an interaction time $T$ with the crystal, in terms of the initial (vacuum) operators,
\begin{eqnarray}
A_{n}^{out}=A_{n}^{in}\cosh[G\sqrt{\lambda_{n}}]
+[B_{n}^{in}]^\dagger \sinh[G\sqrt{\lambda_{n}}], \nonumber\\
B_{n}^{out}=B_{n}^{in}\cosh[G\sqrt{\lambda_{n}}]
+[A_{n}^{in}]^\dagger \sinh[G\sqrt{\lambda_{n}}], \nonumber
\label{Bogol}
\end{eqnarray}
where $A_{n}^{in},B_{n}^{in}$ are the initial (vacuum) photon annihilation operators in the corresponding Schmidt mode~(\ref{ops}) and $G=\Gamma\cdot T$ corresponds to the parametric gain. Also, using the commutation relations one can obtain the Heisenberg equations for the monochromatic-wave operators,
\begin{equation}
 \frac{d a_{\omega_{s}}}{dt}=\Gamma \sum_{n}\sqrt{\lambda_{n}}u_{n}(\omega_{s}) [B_{n}^{out}]^\dagger.
\label{plane}
\end{equation}
The solutions to these equations yield the  output monochromatic-wave operators in terms of  the initial vacuum operators for each frequency from the spectrum. For example, for the signal radiation
\begin{eqnarray}
a^{out}_{\omega_{s}}=a_{\omega_{s}}^{in}+\sum_{n}u_{n}(\omega_{s})[[B^{in}_{n}]^\dagger\sinh(\sqrt{\lambda_{n}}G)\nonumber\\
+A^{in}_{n}(\cosh(\sqrt{\lambda_{n}}G)-1)].
\label{planes}
\end{eqnarray}
In the degenerate case,  the signal and idler photons have the same Schmidt modes, $A_n=B_n$.

 This simple analytical expression allows one to calculate various characteristics of BSV, such as the mean photon number, the variance of the photon number difference in the signal and idler beams, the correlation functions and so on, for different experimental configurations.

\begin{figure}[htb]
\begin{center}
\includegraphics[width=0.5\textwidth]{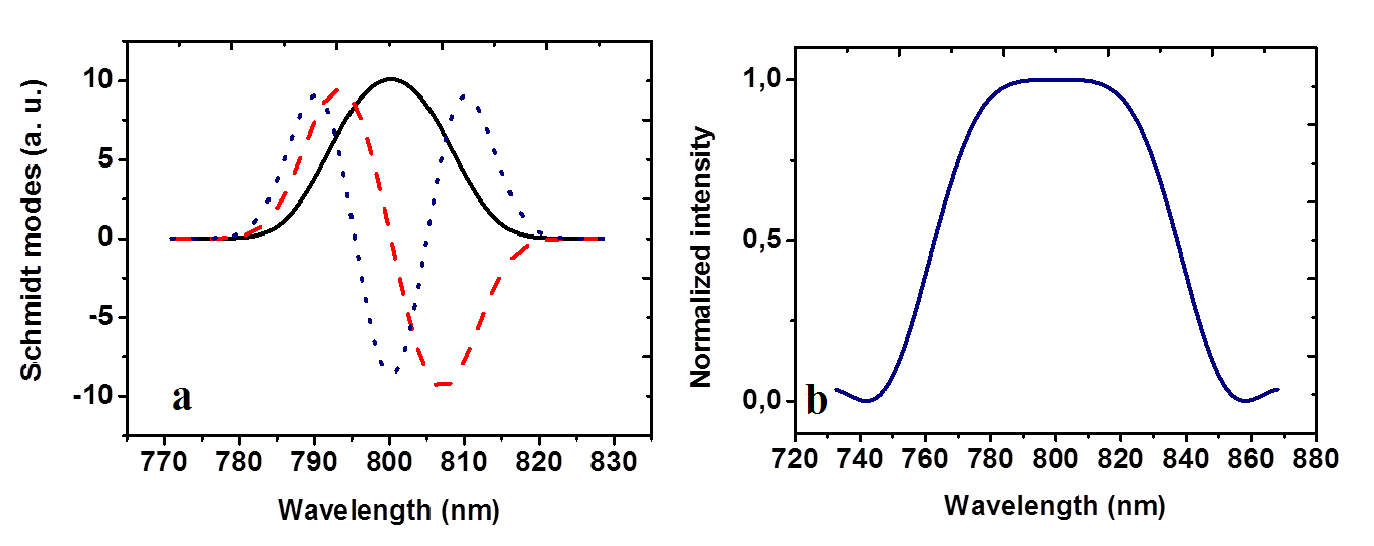}
\end{center}
\caption{For BSV generated in a single 3 mm crystal pumped by 1 ps pulses at the wavelength $\lambda_p=400$ nm: a) the Schmidt modes $u_0$ (black), $u_1$ (red), and $u_2$ (blue); b) the normalized intensity distribution for  parametric gain $G$=1.} \label{fig:Sch_num}
\end{figure}

According to our approach the spectral distribution of the signal beam is given by  the incoherent sum of independent Schmidt modes with the weights $\Lambda_n$,
\begin{equation}
\langle N_s(\omega_s)\rangle=\sum_{n}\vert u_{n}(\omega_s)\vert^{2}\Lambda_n.
\label{num}
\end{equation}
In the simplest case of a single crystal the Schmidt modes are very close to the Hermite functions \cite{Fedorov}. Typical spectral distributions for  three lowest-order Schmidt modes are presented in Fig. \ref{fig:Sch_num}a. The modes contribute independently to the total signal and in the case of a large number of modes give rise to a rather broad spectral distribution (Fig. \ref{fig:Sch_num}b).

\begin{figure}[htb]
\begin{center}
\includegraphics[width=0.5\textwidth]{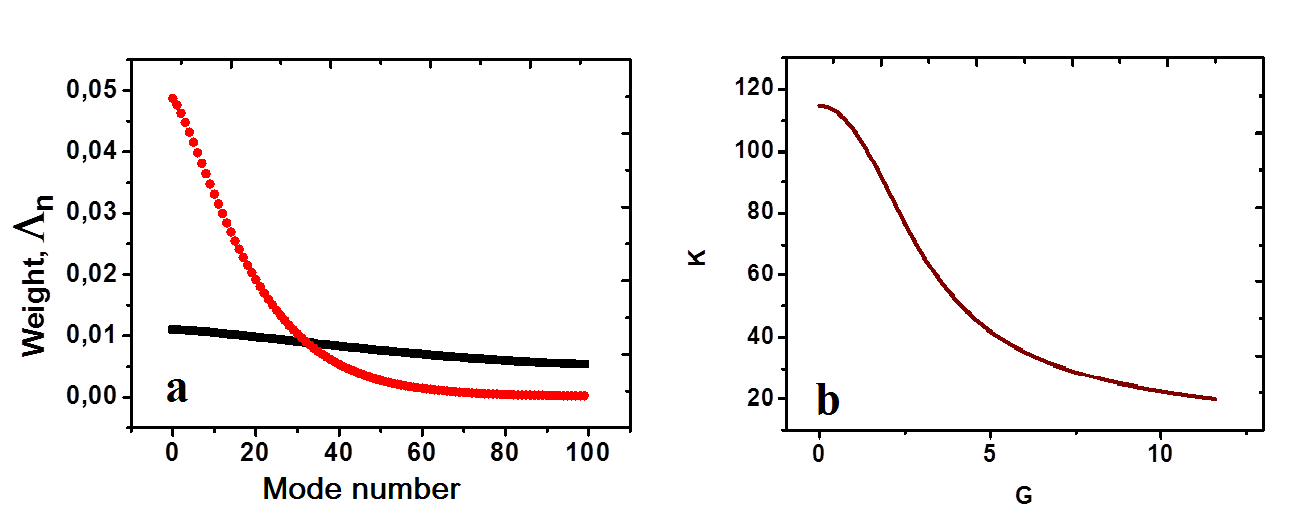}
\end{center}
\caption{Weights of the Schmidt modes for BSV generated under degenerate collinear phase matching in a single BBO crystal of length 3 mm pumped by 1 ps pulses at $400$ nm: (a) the Schmidt eigenvalues for different values of the parametric gain, $G$=1 (black) and $G$=9 (red); (b) the Schmidt number vs  parametric gain. } \label{fig:lambda}
\end{figure}

The weight of each Schmidt mode depends on the parametric gain so that the new Schmidt coefficients $\Lambda_{n}$  determining the contributions  of different modes into the spectral distribution are redistributed. In  the high-gain regime they sufficiently differ from the initial ones $\lambda_{n}$:
\begin{equation}
\Lambda_{n}=\frac{(\sinh[G\sqrt{\lambda_{n}}])^2}{\sum_{n}(\sinh[G\sqrt{\lambda_{n}}])^2}.
\label{new_Schmidt}
\end{equation}
It means that with  increasing the parametric gain the distribution of the Schmidt coefficients becomes  sharper (Fig.~\ref{fig:lambda}a); in other words,  the effective number of modes contributing to the total signal decreases (Fig.~\ref{fig:lambda}b) \cite{Bondani1}.

The effective number of modes is defined by the Schmidt number $K=\dfrac{1}{\sum_n \Lambda_n^2}$ \cite{Bondani1, Timur, Ou} and is reduced with the increase of the parametric gain (Fig.~\ref{fig:lambda}b). It means that in the high-gain limit only the first Schmidt mode will contribute to the total signal and all features of the PDC radiation will be defined by the properties of this Schmidt mode.

 Because the profiles of the Schmidt modes are assumed to not depend on the parametric gain, the model predicts the narrowing of the total spectral width of BSV with the gain increase. While this is not the case for a single crystal, where small spectral broadening can be observed with the gain increase~\cite{Spasibko2012}, the narrowing of the spectrum is indeed observed for a nonlinear interferometer~\cite{exp}, considered in the next section.

\section{Nonlinear interferometer with group-velocity dispersion}

\begin{figure}[htb]
\begin{center}
\includegraphics[width=0.4\textwidth]{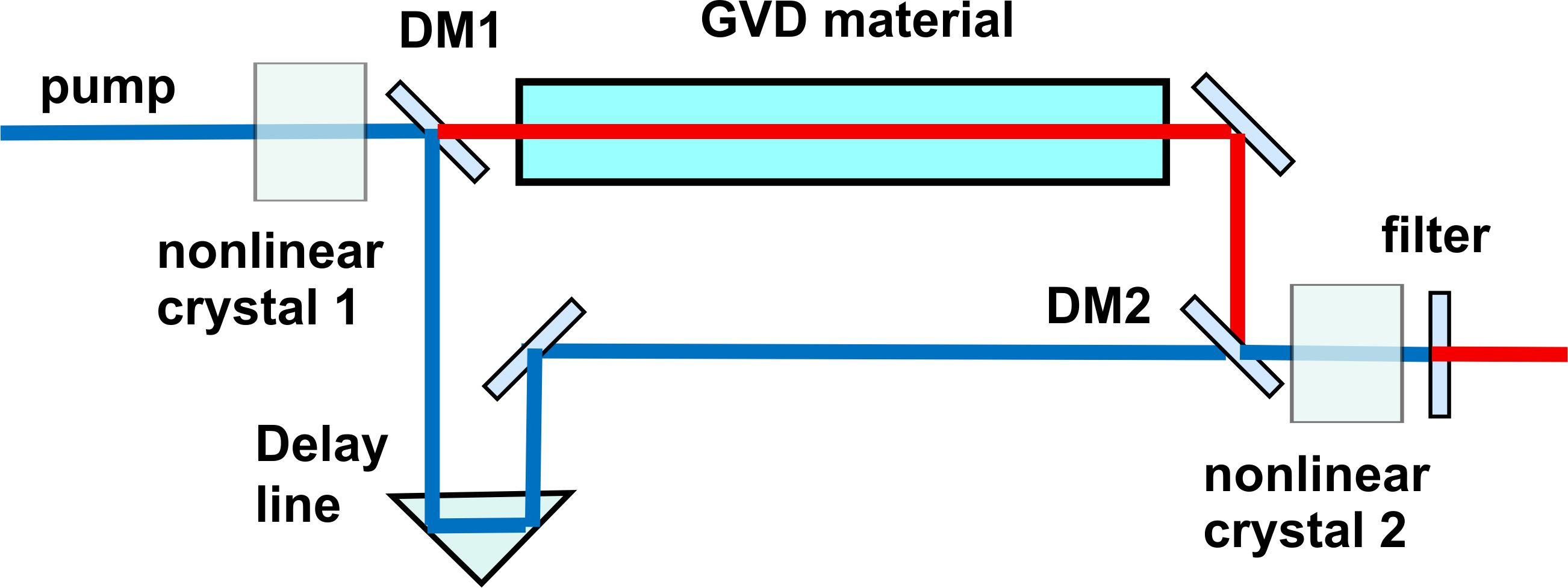}
\end{center}
\caption{An SU(1,1) interferometer with GVD.  BSV is generated in the first nonlinear crystal. The dichroic mirror DM1 separates the pump from the BSV. The BSV propagates through the GVD material while the pump propagates through the air and its time delay can be adjusted.  After the dichroic mirror DM2 the pump and the BSV radiation overlap in the second nonlinear crystal. Finally, the BSV is filtered from the pump.}\label{fig:setup}
\end{figure}

The above-described theoretical approach can be applied to different experimental configurations as long as we can obtain the TPA.  An interesting experimental configuration that allows one to engineer the BSV spectrum and mode content includes two nonlinear crystals separated by a medium with large group velocity dispersion (GVD), a nonlinear SU (1,1) interferometer~\cite{altern,exp,Vered, Shaked} (Fig.~\ref{fig:setup}). In such a configuration, the radiation is down-converted in one nonlinear crystal and can be amplified or de-amplified in the other one, depending on the coherent phase conditions. In the presence of the dispersive medium, BSV generated in the first crystal is spread in time and in addition, acquires a chirp. Its different spectral components propagate inside the GVD medium with their own group velocities. If, in addition, the pump pulse is delayed, only a certain spectral band of the down-converted radiation spectrum will overlap with it in time in the second crystal (Fig.~\ref{fig:setup}) and get amplified there. This way, by changing the time delay of the pump with respect to the PDC radiation one can vary the mode structure of the BSV.

For the two-crystal configuration with the GVD medium (Fig.~\ref{fig:setup}), the TPA (\ref{TPA}) is modified; in addition to the envelope, the modulation term appears \cite{OptLett, Klyshkop}:

\begin{eqnarray}
F(\omega_s,\omega_i)=C \exp\{-\frac{(\omega_s+\omega_i-\omega_p)^2}{2 \Omega^2 }\}\times
\nonumber\\
\mathrm{sinc}(\frac{\Delta k L}{2})
\times\exp\{-i\frac{\Delta k L}{2} \}\times
\nonumber\\
\times\cos\{\frac{\Delta k L+
\Delta k_a d_a+(k_p^a d_0-k_s^g d-k_i^g d)}{2}\} \times
\nonumber\\
\exp\{-i\frac{\Delta k L+
\Delta k_a d_a+(k_p^a d_0-k_s^g d-k_i^g d)}{2} \rbrace
, \ \ \ \ \ \
\label{TPA1}
\end{eqnarray}
where $k_s^g, k_i^g$ are the wavevectors of signal and idler photons in the GVD medium, $d$ is the length of the GVD medium, $d_0$ is the additional pump path length, which can be varied in the experiment, $\Delta k_a=k_{p}^a-k_s^a-k_i^a$ is the wavevector mismatch in the air, with $k_{p,s,i}^a$ being the corresponding wavevectors for the pump, signal, and idler radiation, and $d_a$ is the length of the air gap  where all three beams propagate together. All wavevector mismatches can be calculated directly from the dispersive (Sellmeier) formulas. Let us denote the part of the argument of the cosine function that is due to the dispersive medium and the air gap as $\phi(\omega_s, \omega_i)=(\Delta k_a d_a+(k_p^a d_0-k_s^g d-k_i^g d))/2$.

Depending on this phase and its derivative, the BSV structure can be substantially changed. The derivative of the phase depends on the relation between the group velocities of the pump and the BSV radiation.
Due to varying the additional pump path $d_0$ one can change the phase derivative and satisfy the extremum condition
\begin{equation}
\frac{d \phi}{d \omega_i}=0, \frac{d \phi}{d \omega_s}=0
\label{phase}
\end{equation}
for different frequencies. Such a condition can be considered as a requirement of the group velocity matching between the pump and the chosen BSV frequency. 
In other words, this condition will be fulfilled for the frequency band in the signal radiation that overlaps in time with the pump pulse in the second crystal.

\begin{figure}[htb]
\begin{center}
\includegraphics[width=0.5\textwidth]{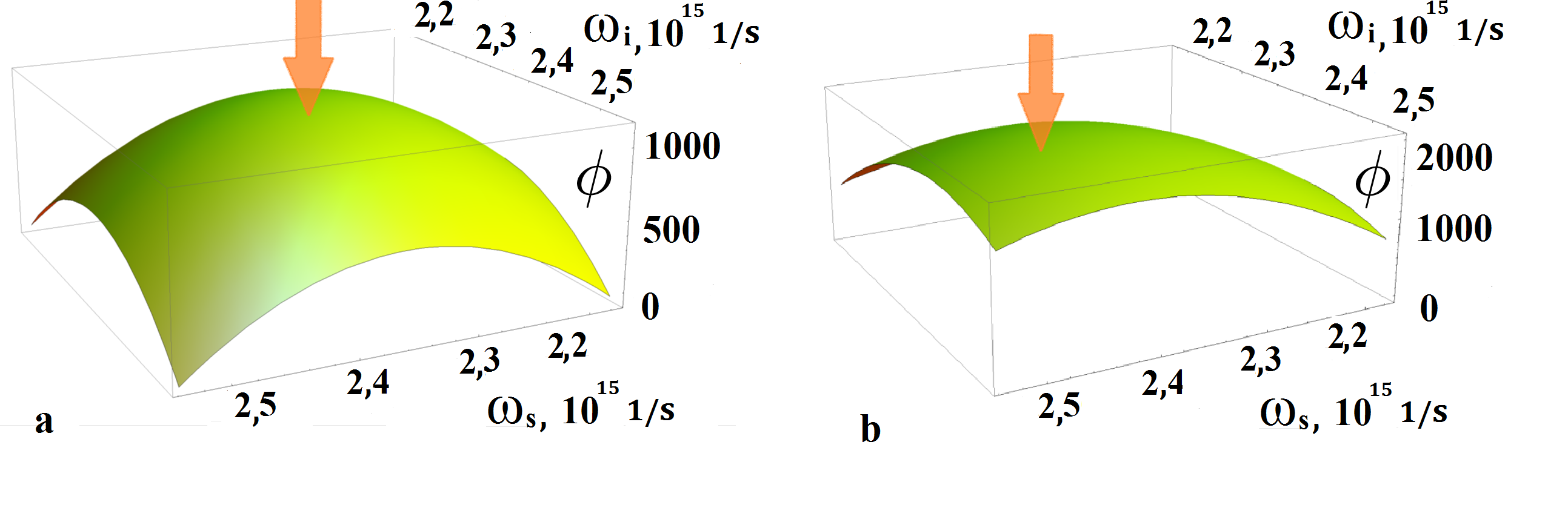}
\end{center}
\caption{The profile of the phase $\phi$  in the cases where the pump pulse overlaps in the second crystal with the  PDC radiation at (a) the degenerate frequancy and (b) a non-degenerate frequency. The orange arrow shows the overlapping frequency. The calculation was done for the case of SF6 Schott glass of length $d=36 cm$ used as the GVD medium. The other parameters are as mentioned above.} \label{fig:Phase}
\end{figure}

 Fig.~\ref{fig:Phase}a shows the phase shape $ \phi$ versus the signal and idler frequencies in the case where condition~(\ref{phase}) is fulfilled for the degenerate frequency $\omega_s=\omega_i=\omega_p/2$. It means that the pump and the PDC radiation at the degenerate frequency (shown by the orange arrow) perfectly overlap in time in the second nonlinear crystal.

If the pump pulse overlaps in time in the second crystal with the  PDC  radiation for a certain non-degenerate frequency, the extremum condition (\ref{phase}) will be fulfilled for this frequency. Fig.~\ref{fig:Phase}b shows the phase profile in such a case, the orange arrow shows the chosen frequency.

Another factor affecting the BSV features is the total value of the phase $\phi$. Depending on $\phi$, the radiation from the first crystal is amplified or deamplified in the second one. For a fixed frequency, this can be changed by slightly varying the  pump path between the GVD medium and the second crystal. Indeed, this almost does not affect the group delay but changes sufficiently the total phase. Thereby by changing the experimental parameters we can obtain different shapes of the TPA. And as long as we know the TPA for the chosen experimental configuration, we can provide its Schmidt decomposition and apply the theoretical approach described above.

\begin{figure}[htb]
\begin{center}
\includegraphics[width=0.4\textwidth]{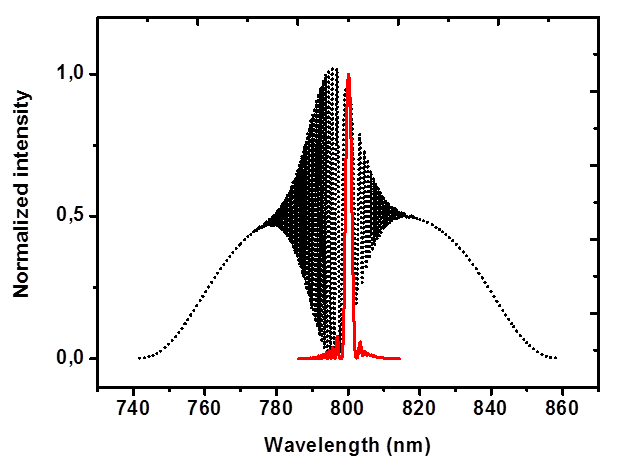}
\end{center}
\caption{Normalized intensity distribution for an SU (1,1) interferometer with 36 cm of SF6 glass. The pump pulse overlaps in the second crystal with the frequency-degenerate part of the PDC radiation. Black dotted line: parametric gain $G=1$, red solid line: parametric gain $G=13$.} \label{fig:Without_delay}
\end{figure}

To investigate the effect of the GVD medium on the BSV structure we consider $d=36$ cm of highly dispersive glass (SF6, $k''=199.01 \,\rm{fs}^2/\rm{mm}$). First, we chose the additional pump path and the air gap between the GVD medium and the second crystal so that both conditions, for overlapping between the pump and signal pulse in the second crystal (\ref{phase}) and for the amplification ($\phi=0$), are satisfied for the degenerate frequency. The intensity distribution calculated for the case of low parametric gain is shown in Fig.~\ref{fig:Without_delay} by a black curve. It has a rather broad envelope with fast interference oscillations at the center. The interference fringes as well as the broad spectrum profile indicate the multimode structure of the down-converted radiation. As the parametric gain increases, the number of modes is reduced, and the spectrum gets narrower. At high gain (Fig.~\ref{fig:Without_delay}, red curve), the intensity distribution shrinks drastically compared to the low-gain regime and, as we will show further, the radiation becomes nearly single-mode in this case.

 One can see a similarity between the spectra shown in Fig.~\ref{fig:Without_delay} for lower and higher gain values and the results of Ref.~\cite{altern} and Ref.~\cite{exp}, respectively.

\begin{figure}[htb]
\begin{center}
\includegraphics[width=0.5\textwidth]{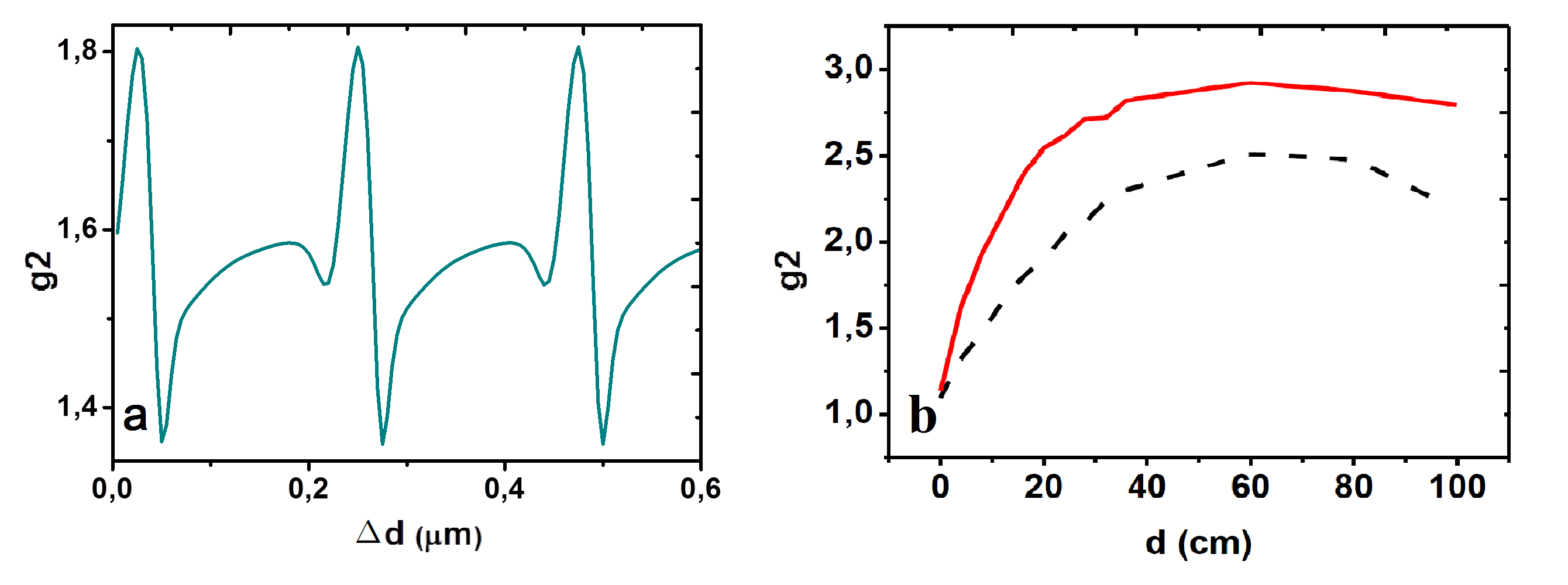}
\end{center}
\caption{Normalized second-order intensity correlation function at the output of the SU(1,1) interferometer calculated versus the length of the SF6 glass for the same configuration as in Fig.~\ref{fig:Without_delay}: (a)  for the length of the GVD medium $d=\tilde{d}+\Delta d$, $\tilde{d}=5\ \rm{cm}$, the parametric gain $G=13$ and (b) under conditions (\ref{phase}) and $\phi$=0 for different parametric gain values: $G$=8.5 (black dashed line), $G$=13 (red solid line).}  \label{fig:g2}
\end{figure}

The effective number $K$ of the Schmidt modes is given by the second-order normalized intensity correlation function for the integral spectrum, $g^{(2)}=1+2/K$ for the degenerate case. The calculated dependence of $g^{(2)}$ on the length of the SF6 glass is shown in Fig.~\ref{fig:g2}a.
One can observe  a fast modulation, on the micrometer scale, caused by the variation of $\phi$ and the resulting oscillations from amplification to deamplification.
Even a very small change in the GVD medium length makes the total phase significantly different,  which results in the strong non-monotonic dependence of $g^{(2)}$ on the medium length. The oscillation period at the degenerate wavelength can be calculated from the equation $(k^{g}_s|_{\frac{\omega_p}{2}}+k^{g}_i|_{\frac{\omega_p}{2}})d=\pi$ and is $0.224 \mu$m.  The sharp peaks in the correlation function are much narrower than oscillations in a conventional interferometer and indicate its phase supersensitive features~\cite{Manceau}. The same behavior had been observed in the case of an interferometer with the air gap~\cite{OptLett}.

These fast oscillations can be eliminated by providing constructive interference for a given wavelength through phase locking the interferometer. Under such a condition, the correlation function grows monotonically with the increase in the GVD medium length (Fig.~\ref{fig:g2} b), achieves its maximal value and then decreases due to the contribution of higher-order Schmidt modes.

From Fig.~\ref{fig:g2}b, it is also clear that with increasing the parametric gain, the maximal value of $g^{(2)}$ goes up. For the gain $G=13$, it achieves $g^{(2)}=3$, which corresponds to the case of a single temporal mode. Thus, by choosing appropriate experimental parameters, namely, a sufficiently long GVD medium and a sufficiently high pump power at the same time, one can achieve BSV with a single frequency mode populated with a huge number of photons.

It is worth noting that in the case of a single crystal, the number of modes is also reduced with increasing the parametric gain but due to the initial huge number of modes, the single-mode regime is not achievable for reasonable pump intensities. Using the GVD medium is a much more efficient instrument for reducing the number of modes.

\begin{figure}[htb]
\begin{center}
\includegraphics[width=0.4\textwidth]{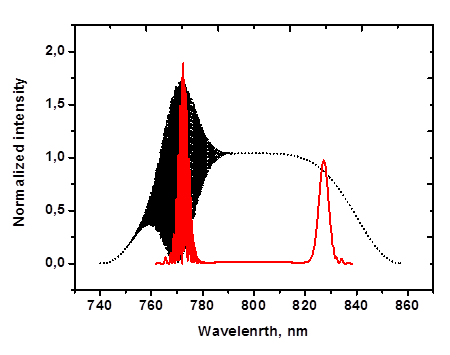}
\end{center}
\caption{Normalized intensity distribution for an SU (1,1) interferometer with  36 cm of SF6 glass. The pump pulse overlaps in the second crystal with the PDC radiation at non-degenerate wavelength $827$ nm. Black dotted line: parametric gain $G=1$, red solid line: parametric gain $G=13$.} \label{fig:With_delay}
\end{figure}

If, by changing the pump path, one makes the pump pulse overlap in the second crystal with the PDC  radiation at a certain non-degenerate frequency, the extremum condition will be fulfilled for the conjugated frequency (Fig.~\ref{fig:Phase}b). In this case, the spectral intensity distribution at low gain is also broad, but the interference fringes will be only observed in the conjugated frequency range (Fig.~\ref{fig:With_delay}). This is a manifestation of the induced coherence effect \cite{induced}: in order to observe interference fringes in the signal radiation from the first and the second crystals, the idler radiation from both crystals should be indistinguishable (in our case, overlap in time).

However, as the parametric gain increases, the situation changes dramatically: instead of a single broad peak, two separated peaks appear, as for non-degenerate (`two-color') BSV generation~\cite{Agafonov}. One of these peaks is observed at the frequency satisfying condition (\ref{phase}), the other one at the conjugated frequency. The second peak has interference structure while the first one is smooth. Thus with increasing the parametric gain the frequency spectrum  gets narrower. If in this case the pump delay is changed this process becomes tunable.

\section{SCHMIDT MODES OF TWO-COLOUR BSV.}
\begin{figure}[htb]
\begin{center}
\includegraphics[width=0.5\textwidth]{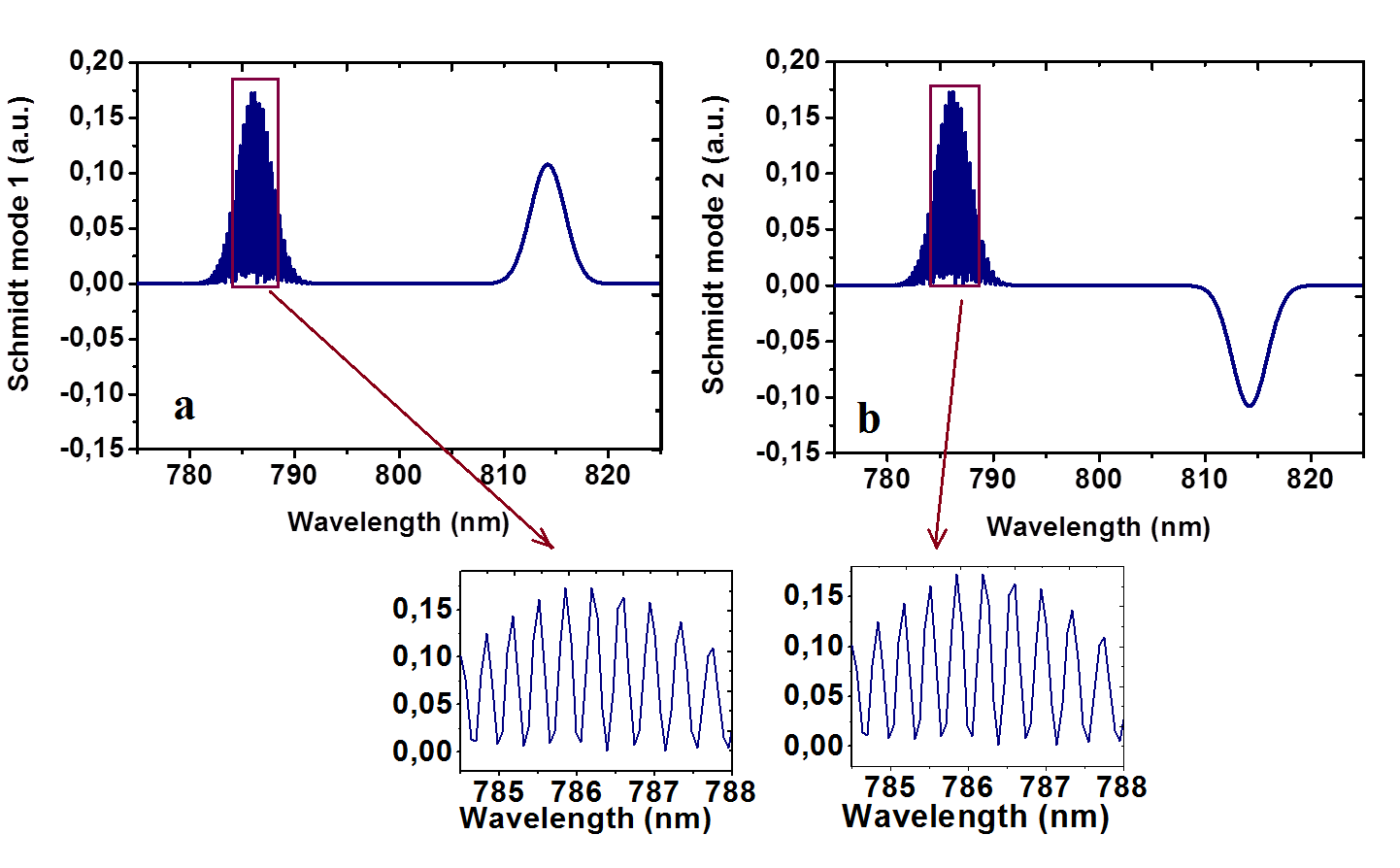}
\end{center}
\caption{The Schmidt modes $u_0,u_1$ for the case shown in Fig.~\ref{fig:With_delay}, where, due to the high gain and a delay introduced in the interferometer, BSV has only two modes and its spectrum has a double-peak structure: (a) mode $u_0$ has a symmetric envelope, b) mode $u_1$ has an antisymmetric envelope. The insets show the left peak of each mode with a better resolution. }
\label{fig:Schmidt_mode}
\end{figure}

The first and second Schmidt modes in the case of two-color PDC (Fig.~\ref{fig:With_delay}) are shown in Fig.~\ref{fig:Schmidt_mode}a,b. One can see that they have a double-peak structure. These modes  $u_0,u_1$ have the same eigenvalues in the Schmidt decomposition and the same intensity profiles, but different symmetry: for the first Schmidt mode, the envelope is symmetric with respect to the degenerate frequency, for the second one it is antisymmetric. It means that even at a sufficiently high gain, BSV will be characterized by two-mode structure, each mode having a double peak profile.

 Because the exact phase matching is achieved for the degenerate wavelength, the idler Schmidt modes have the same shape as the signal ones, i.e. $A_n=B_n$ and the diagonalized Hamiltonian (\ref{Hams}) can be rewritten in the form
\begin{equation}
H=i\hbar\Gamma\sum_{n} \sqrt{\lambda_{n}}(A_{n}^{\dagger 2} -A_{n}^2) .
\label{Hamq}
\end{equation}
For each mode $u_n$, as the ones shown in Fig.~\ref{fig:Schmidt_mode},  one could observe quadrature squeezing; however such an experiment is rather difficult because the local oscillator should be prepared with the same complicated profile as shown in Fig.~\ref{fig:Schmidt_mode} a or b.

On the other hand, instead of these 'odd' and 'even' Schmidt modes $u_n,u_{n+1}$, with the photon creation operators $A^\dagger_n,A^\dagger_{n+1}$, one can pass to their superpositions, corresponding to the operators $C_{n,n+1}^\dagger=\frac{1}{2}(A_n^\dagger+A^\dagger_{n+1})$ and $D_{n,n+1}^\dagger=\frac{1}{2}(A_n^\dagger-A^\dagger_{n+1})$.  While the first one will have the shape of a single modulated peak (the  left-hand peak in Fig.~\ref{fig:Schmidt_mode}a), the second mode will have the shape of a smooth peak (the  right-hand peak in Fig.~\ref{fig:Schmidt_mode}a). For these modes it is possible to observe the twin-beam squeezing which can be  measured using a spectral filter for selecting $C_{n,n+1}$ and $D_{n,n+1}$ modes. The quantitative characteristic of the twin-beam squeezing is the noise reduction factor (NRF),
\begin{equation}
NRF=\frac{\langle(N_s-N_i)^2\rangle-\langle N_s-N_i\rangle^2}{\langle N_s\rangle+\langle N_i\rangle},
\label{NRF}
\end{equation}
where $\langle N_s\rangle$ and $\langle N_i\rangle$ are the integrated numbers of photons in the signal and idler beams.

The condition $NRF<1$ is a signature of twin-beam squeezing. For the double-peak structure of Fig.~\ref{fig:With_delay}, the left peak corresponding to the signal beam and the right peak to the idler one, calculation yields $NRF=10^{-8}$ due to accuracy. This demonstrates an almost perfect twin-beam squeezing, the difference from zero caused by the oscillating structure of the left peak.

\section{CONCLUSION}

We have presented a fully analytical approach to the description of the frequency properties of BSV, based on the model of independent Schmidt modes. Within this approach, we have described the operation of an SU(1,1) interferometer with a dispersive material and its effect on the Schmidt-mode structure of the generated BSV. We have shown that with the transition from low to high parametric gain, the interference structure in the spectrum is replaced by a single- or a double-peak structure, depending on the path length difference in the interferometer. In the second case, the Schmidt modes  also have a double-peak structure. By appropriately shaping the local oscillator, one can observe quadrature squeezing for each of the 'double-peak' mode. This, however, is difficult due to the modulation of one of the peaks caused by the 'induced coherence' effect. Alternatively, and in a simpler way, one can observe twin-beam squeezing by selecting the two peaks in the spectrum separately and registering their variance of the intensity difference.

\section{ACKNOWLEDGMENTS}

We acknowledge the financial support of the joint DFG-RFBR project CH 1591/2-1, 16-52-12031 NNIOa and the Deutsche Forschungsgemeinschaft (DFG)
via TRR~142/1, project C02. We acknowledge partial financial support
of the Russian Foundation for Basic Research, Grant No 18-02-00645a.  S.L. and R.W.B. acknowledge support from the Canada Excellence Research Chairs program and from the Max Planck Centre for Extreme and Quantum Photonics. P.~R.~Sh. thanks the state of North Rhine-Westphalia for support by the
{\it Landesprogramm f{\"u}r geschlechtergerechte Hochschulen}.


\begin{thebibliography}{99}



\bibitem{Jedr} O. Jedrkiewicz, Y.-K. Jiang, E. Brambilla, A. Gatti, M. Bache, L. A. Lugiato, and P. Di Trapani, Physical Review Letters \textbf{93}, 243601 (2004).

\bibitem{Bondani} M. Bondani, A. Allevi, G. Zambra, M. G. A. Paris, and A. Andreoni, Physical Review A \textbf{76}, 013833 (2007).

\bibitem{Brida} G. Brida, L. Caspani, A. Gatti, M. Genovese, A. Meda, and I. R. Berchera, Physical Review Letters \textbf{102}, 213602 (2009).

\bibitem{Agafonov}I. N. Agafonov, M. V. Chekhova, and G. Leuchs, Phys. Rev. A \textbf{82}, 011801 (2010).

\bibitem{Lett} Neil Corzo et al., Optics Express \textbf{19}, 21358 (2011).

\bibitem{Shaked2017}Y.~Shaked, Y.~Michael, R.~Vered, L.~Bello, M.~Rosenbluh and A.~Pe’ers, arXiv:1701.07948v3 [physics.optics] (2017).

\bibitem{nonsep}T. Sh. Iskhakov, I. N. Agafonov, M.V. Chekhova, and G. Leuchs, Phys. Rev. Lett. \textbf{109}, 150502 (2012).

\bibitem{Brida2010}  G. Brida, M. Genovese, and I. R. Berchera, Nature Photonics \textbf{4}, 227 (2010).

\bibitem{Boyer} V.~Boyer, A.~M.~Marino, R.~C.~Pooser, P.~D.~Lett, Science \textbf{321}, 544 (2008).

\bibitem{Bondani_narrow} Alessia Allevi and Maria Bondani, J. Opt \textbf{19}, 064001 (2017).

\bibitem{Brida_OpEx2010}  G.~Brida, I.~P.~Degiovanni, M.~Genovese, M.~L.~Rastello, and I. R.-Berchera, Optics Express \textbf{18}, 20572 (2010).

\bibitem{Manceau} M.~Manceau, G. Leuchs, F.~Ya.~Khalili, and M.V. Chekhova, Phys. Rev. Lett. \textbf{119}, 223604 (2017).

\bibitem{Boyer1}V. Boyer, A.M. Marino, and P. D. Lett, PRL \textbf{100}, 143601 (2008).

\bibitem{OptLett} A. M. P´erez, T. Sh. Iskhakov, P. Sharapova, S. Lemieux, O. V. Tikhonova,M. V. Chekhova, and G. Leuchs, Opt. Lett. \textbf{39}, 2403 (2014).

\bibitem{Braunstein} S.~L.~Braunstein, PRA \textbf{71}, 055801 (2005).

\bibitem{Boyd}R.~S.~Bennink and R.~W.~Boyd, PRA \textbf{66}, 053815 (2002).

\bibitem{Wasilewski} W.~Wasilewski, A.~I.~Lvovsky, K.~Banaszek, and C.~Radzewicz, PRA \textbf{73}, 063819 (2006).

\bibitem{Dayan} Barak Dayan, PRA \textbf{76}, 043813 (2007).

\bibitem{Christ} Andreas Christ, Benjamin Brecht, Wolfgang Mauerer, and Christine Silberhorn, New Journ. of Phys. \textbf{15}, 053038 (2013).

\bibitem{Eckstein} Andreas Eckstein, Benjamin Brecht, and Christine Silberhorn, Opt. Express \textbf{19}, 13770 (2011).

\bibitem{Vered}Rafi Z. Vered, Yaakov Shaked, Yelena Ben-Or, Michael Rosenbluh, and Avi Pe’er, Phys. Rev. Lett.  \textbf{114}, 063902 (2015).

\bibitem{Shaked}Y. Shaked, R. Pomerantz, R. Z. Vered, and A. Peer, New J. Phys. \textbf{16}, 053012 (2014).

\bibitem{altern}T. Sh. Iskhakov, S. Lemieux, A. M. Perez, R. W. Boyd, M. V. Chekhova, and G. Leuchs, Journal of Modern Optics \textbf{63}, 64 (2016).

\bibitem{exp}Samuel Lemieux, Mathieu Manceau, Polina R. Sharapova, Olga V. Tikhonova, Robert W. Boyd, Gerd Leuchs and Maria V. Chekhova, Phys. Rev. Lett.  \textbf{117}, 183601 (2016).

\bibitem{Yurke}B. Yurke, S.L. McCall, and J.R. Klauder, Phys. Rev. A \textbf{33}, 4033 (1986).

\bibitem{Jing} J. Jing, C. Liu, Z. Zhou, Z. Y. Ou, and W. Zhang, Appl. Phys. Lett. \textbf{99}, 011110 (2011).

\bibitem{Hudelist} F. Hudelist, J. Kong, C. Liu, J. Jing, Z. Y. Ou, and W. Zhang, Nat. Commun \textbf{5}, 3049 (2014).

\bibitem{Klyshkoint} D. N. Klyshko, Zh. Eksp. Teor. Fiz. \textbf{104}, 2676-268 (1993).

\bibitem{PRA} P. Sharapova, A. M. Perez, O. V. Tikhonova, and M. V. Chekhova, Physical Review A \textbf{91}, 043816 (2015).

\bibitem{Spasibko2012}  K.~Yu.~Spasibko, T.~Sh.~Iskhakov, and M.~V.~Chekhova, Opt. Express \textbf{20}, 7507 (2012).

\bibitem{Klyshkop}D. N. Klyshko, Photons and Nonlinear Optics. Gordon and Breach Science Publishers (1988).

\bibitem{Sellm} V.G. Dmitriev, G.G. Gurzadyan, D.N. Nikogosyan, Handbook of Nonlinear Optical Crystals. Springer, 1999.

\bibitem{Law} C. K. Law and J. H. Eberly, Phys. Rev. Lett. \textbf{84}, 5304 (2000).

\bibitem{Fedorov} M. V. Fedorov, Yu. M. Mikhailova, and P. A. Volkov, J. Phys. B: At. Mol. Opt. \textbf{42}, 175503 (2009).

\bibitem{FedorovPRA} M. V. Fedorov, M. A. Efremov, P. A. Volkov, E. V. Moreva, S. S. Straupe, S. P. Kulik, Phys. Rev. A \textbf{77}, 032236 (2008).

\bibitem{Bondani1} A. Allevi, O. Jedrkiewicz, E. Brambilla, A. Gatti, J. Perina, Jr. O. Haderka, and M. Bondani, PRA, \textbf{90}, 063812 (2014).

\bibitem{Timur} I. V. Dyakonov, P. R. Sharapova, T. Sh. Iskhakov, G. Leuchs, Laser. Phys. Lett.,\textbf{12}, 065202 (2015).

\bibitem{Ou}X. Guo, N. Liu, X. Li, Z. Y. Ou, Opt. Express \textbf{23}, 29369 (2016).

\bibitem{review} M. V. Chekhova and Z. Y. Ou, Advances in Optics and Photonics \textbf{8}, 104 (2016).

\bibitem{induced} X. Y. Zou, L. J. Wang, and L. Mandel, Phys. Rev. Lett. \textbf{67}, 3 (1991).















\end{thebibliography}
\end{document}